\documentclass[]{elsarticle}

\usepackage{graphicx}
\usepackage{amsmath}
\usepackage{amssymb}
\usepackage{hyperref}
\usepackage{color}

\begin{document}
\sloppy
\title{A duality web of linear quivers}

\author[fb]{Frederic Br\"unner}
\ead{bruenner@hep.itp.tuwien.ac.at}
\address[fb]{Institute for Theoretical Physics, Vienna University of Technology\\
        Wiedner Hauptstra\ss e 8-10, A-1040 Vienna, Austria}

\author[vs]{Vyacheslav P. Spiridonov}
\ead{spiridon@theor.jinr.ru}
\address[vs]{Laboratory of Theoretical Physics, JINR, Dubna, Moscow region, 141980, Russia}

\begin{abstract}
We show that applying the Bailey lemma to elliptic hypergeometric integrals on the
$A_n$ root system leads to a large web of dualities for $\mathcal{N} = 1$ supersymmetric linear
quiver theories. The superconformal index of Seiberg's SQCD with $SU(N_c)$ gauge group and
$SU(N_f)\times SU(N_f)\times U(1)$ flavour symmetry is equal to that of $N_f-N_c-1$ distinct
linear quivers. Seiberg duality further enlarges this web by adding new quivers.
In particular, both interacting electric and magnetic theories with arbitrary $N_c$
and $N_f$ can be constructed by quivering an $s$-confining theory with $N_f=N_c+1$.
\end{abstract}

\maketitle

Supersymmetric gauge theories are a highly active subject of
study and many discoveries were made in this field in the
past decades. One particularly interesting phenomenon
is duality: for certain strongly coupled supersymmetric
quantum field theories, there exist weakly coupled dual
theories that describe the same physical system in terms
of different degrees of freedom. A famous example is Seiberg duality \cite{seiberg} for $\mathcal{N}=1$ supersymmetric quantum chromodynamics (SQCD), where two dual
theories, referred to as electric and magnetic, flow to the same infrared (IR) theory. While such dualities are hard to prove, supersymmetric
theories allow for the definition of observables that
are independent of the description, i.e. they should yield the same result on both sides of the duality.
One such quantity is the superconformal index (SCI)
\cite{KMMR,R}, which counts the number of BPS states of a given theory. It turns out that SCIs are related to elliptic 
hypergeometric functions, which have also found many other 
applications in physics.

A long hunt for the most general possible exactly solvable model of
quantum mechanics has led to the discovery of
elliptic hypergeometric integrals forming a new class of transcendental
special functions \cite{spi}. In the first physical
setting these integrals served either as a normalization condition of
particular eigenfunctions or as eigenfunctions of the
Hamiltonian of an integrable Calogero-Sutherland type model \cite{spi2007}.
The Bailey lemma for such integrals \cite{spi2004} appeared to define the star-triangle relation
associated with quantum spin chains \cite{DS}. However, a major
physical application was found by Dolan and Osborn \cite{DO} who
showed that certain elliptic hypergeometric integrals are identical to
SCIs  of $4d$ supersymmetric field theories and that Seiberg duality can be understood in terms of symmetries of such integrals. In \cite{SV1}, many explicit examples were studied.  In the present work, we describe a web of dualities that can be constructed using the Bailey lemma of \cite{spi2004} and \cite{SW}. Starting
from a known elliptic beta integral on the $A_n$ root system \cite{spi03} that is identified with the
star-triangle relation, one gets an algorithm for constructing an infinite
chain of symmetry transformations for elliptic hypergeometric integrals. The emerging integrals can be interpreted as the SCIs  of linear quiver gauge theories, a possibility that was already mentioned in \cite{SV1}.

Quiver gauge theories are theories with product gauge groups that
arise as world volume theories of branes placed on singular spaces or
from brane intersections \cite{DM,HW,GK}. Their field content can be depictured by so-called
quiver diagrams; all new theories discussed in this article are of this type. Note that while the quivers we discuss are also linear like those described in \cite{BB}, field content and flavour symmetries are different.

\begin{figure*}[t!]
\centering
\def\svgwidth{450pt}

\begingroup%
  \makeatletter%
  \providecommand\color[2][]{%
    \errmessage{(Inkscape) Color is used for the text in Inkscape, but the package 'color.sty' is not loaded}%
    \renewcommand\color[2][]{}%
  }%
  \providecommand\transparent[1]{%
    \errmessage{(Inkscape) Transparency is used (non-zero) for the text in Inkscape, but the package 'transparent.sty' is not loaded}%
    \renewcommand\transparent[1]{}%
  }%
  \providecommand\rotatebox[2]{#2}%
  \ifx\svgwidth\undefined%
    \setlength{\unitlength}{1024.94503209bp}%
    \ifx\svgscale\undefined%
      \relax%
    \else%
      \setlength{\unitlength}{\unitlength * \real{\svgscale}}%
    \fi%
  \else%
    \setlength{\unitlength}{\svgwidth}%
  \fi%
  \global\let\svgwidth\undefined%
  \global\let\svgscale\undefined%
  \makeatother%
  \begin{picture}(1,0.28192802)%
    \put(0,0){\includegraphics[width=\unitlength]{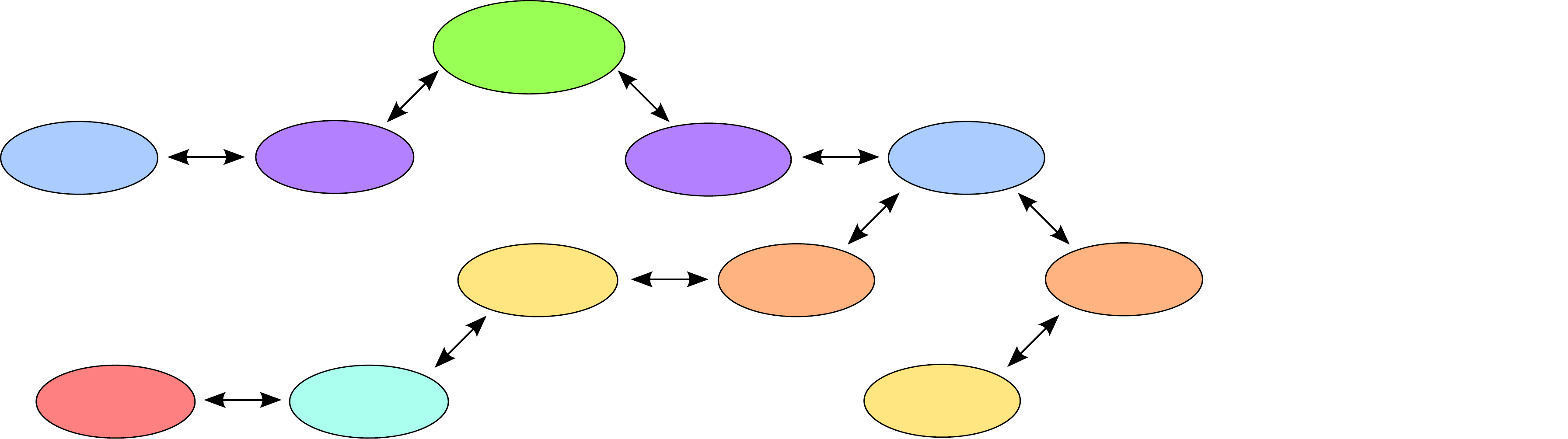}}%
    \put(0.31899153,0.24347951){\color[rgb]{0,0,0}\makebox(0,0)[lb]{\smash{\scalebox{.6}{$\mathrm{SU(3)}$}}}}%
    \put(0.43527629,0.17323184){\color[rgb]{0,0,0}\makebox(0,0)[lb]{\smash{\scalebox{.6}{$\mathrm{SU(3)^2}$}}}}%
    \put(0.19510679,0.17323184){\color[rgb]{0,0,0}\makebox(0,0)[lb]{\smash{\scalebox{.6}{$\mathrm{SU(3)^3}$}}}}%
    \put(0.03200085,0.17323184){\color[rgb]{0,0,0}\makebox(0,0)[lb]{\smash{\scalebox{.6}{$\mathrm{SU(3)^2}$}}}}%
    \put(0.57438222,0.17323184){\color[rgb]{0,0,0}\makebox(0,0)[lb]{\smash{\scalebox{.6}{$\mathrm{SU(3)\times SU(2)}$}}}}%
    \put(0.67145107,0.09517887){\color[rgb]{0,0,0}\makebox(0,0)[lb]{\smash{\scalebox{.6}{$\mathrm{SU(3)\times SU(2)^3}$}}}}%
    \put(0.55517694,0.0171259){\color[rgb]{0,0,0}\makebox(0,0)[lb]{\smash{\scalebox{.6}{$\mathrm{SU(3)\times SU(2)^2}$}}}}%
    \put(0.46170806,0.09517887){\color[rgb]{0,0,0}\makebox(0,0)[lb]{\smash{\scalebox{.6}{$\mathrm{SU(3)\times SU(2)^2}$}}}}%
    \put(0.29800214,0.09517887){\color[rgb]{0,0,0}\makebox(0,0)[lb]{\smash{\scalebox{.6}{$\mathrm{SU(3)\times SU(2)^2}$}}}}%
    \put(0.19122798,0.0171259){\color[rgb]{0,0,0}\makebox(0,0)[lb]{\smash{\scalebox{.6}{$\mathrm{SU(3)\times SU(2)^3}$}}}}%
     \put(0.02822798,0.0171259){\color[rgb]{0,0,0}\makebox(0,0)[lb]{\smash{\scalebox{.6}{$\mathrm{SU(3)\times SU(2)^2}$}}}}
    \put(0.4254692,0.21927602){\color[rgb]{0,0,0}\makebox(0,0)[lb]{\smash{\tiny{$Q$}}}}%
    \put(0.2383103,0.21927602){\color[rgb]{0,0,0}\makebox(0,0)[lb]{\smash{\tiny{$Q$}}}}%
    \put(0.68443337,0.14141776){\color[rgb]{0,0,0}\makebox(0,0)[lb]{\smash{\tiny{$Q$}}}}%
    \put(0.52930098,0.14141776){\color[rgb]{0,0,0}\makebox(0,0)[lb]{\smash{\tiny{$Q$}}}}%
    \put(0.26572619,0.06117008){\color[rgb]{0,0,0}\makebox(0,0)[lb]{\smash{\tiny{$Q$}}}}%
    \put(0.6363804,0.06117008){\color[rgb]{0,0,0}\makebox(0,0)[lb]{\smash{\tiny{$S$}}}}%
    \put(0.53351684,0.19166542){\color[rgb]{0,0,0}\makebox(0,0)[lb]{\smash{\tiny{$S$}}}}%
    \put(0.42424799,0.11361245){\color[rgb]{0,0,0}\makebox(0,0)[lb]{\smash{\tiny{$S$}}}}%
    \put(0.12694141,0.19166542){\color[rgb]{0,0,0}\makebox(0,0)[lb]{\smash{\tiny{$S$}}}}%
    \put(0.1519141,0.03906542){\color[rgb]{0,0,0}\makebox(0,0)[lb]{\smash{\tiny{$S$}}}}%
  \end{picture}%
\endgroup%

\caption{The duality web corresponding to the electric part of SQCD for $N_c=3$ and $N_f=6$. $Q$ denotes a duality obtained from Eq. (\ref{equiver}) and $S$ denotes Seiberg duality of Eq. (\ref{seiberg}). In total, there are ten distinct quiver gauge theories dual to the original theory.}
\label{network}
\end{figure*}
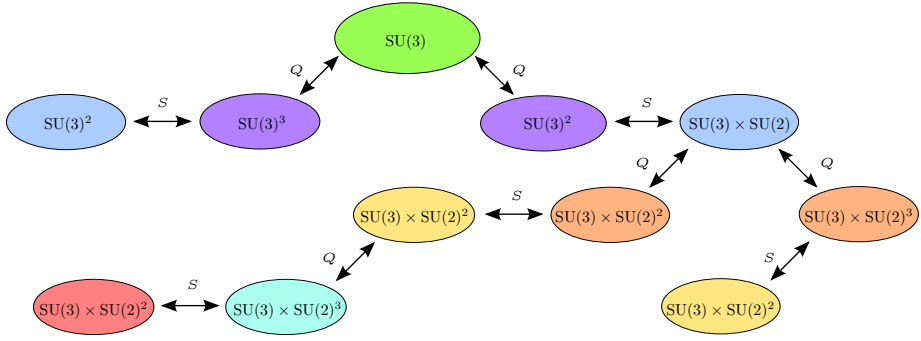

This letter is dedicated to applying an integral extension of the standard
Bailey chains techniques \cite{AAR} to SCIs.  We identify the star-triangle
relation (a variant of the Yang-Baxter
equation) with an elliptic hypergeometric integral on the
$A_n$ root system that corresponds to the superconformal
index of an $s$-confining $\mathcal{N} = 1$ $SU(N_c)$ gauge theory.
The main result of our calculation is that the SCI  of SQCD with $SU(N_c)$ gauge group and $SU(N_f)\times SU(N_f)\times U(1)$ flavour symmetry is equal to that of $N_f-N_c-1$ distinct linear quivers. Seiberg duality leads to magnetic partners for these quivers, some of which are again dual to yet other quivers.
Corresponding indices can also be shown to be equivalent to that of another set of quivers. In total, this leads to a very large duality web, composed of Seiberg and Bailey lemma dualities. An example of such a web corresponding to the electric SQCD with $N_c=3$ and $N_f=6$ is illustrated in Fig. \ref{network}. Another nontrivial consequence is that indices of both electric and magnetic interacting theories can be constructed from a simple $s$-confining theory.

The SCI of $\mathcal{N}=1$ theories is defined as
\begin{equation}
\mathcal{I}=\mathrm{Tr}(-1)^{\mathcal{F}}e^{-\beta H}p^{\frac{R}{2}+J_R+J_L}q^{\frac{R}{2}+J_R-J_L}
\prod_{i} z_i^{G_i}\prod_{j} y_j^{F_j}\!\!,
\end{equation}
\noindent where $\mathcal{F}$ is the fermion number, $R$ is the $R$-charge, $J_L$ and $J_R$ are the Cartan generators of the rotation group $SU(2)_L\times SU(2)_R$, and $G_i$ and $F_j$ are maximal torus generators of the gauge and flavour groups. The theory is assumed to be compactified on a spatial three-sphere, hence the name ``sphere index''. As shown in \cite{ACM} (see also \cite{ACDKLM} and \cite{BBK}), in this case the SCI is proportional to the partition function of the theory, where $p$ and $q$ are variables of the three-sphere metric and the parameters $y_j$ are interpreted as mean values of the background gauge fields of the flavour group. The index only receives contributions from states with $H=E-2J_L-\frac{3}{2}R=0$, $E$ being the energy, and is independent of the chemical potential $\beta$. In order to obtain a gauge invariant expression, an integral over the gauge group is performed, which gives the explicit expression
\begin{equation}\label{index}
\mathcal{I}(p,q,y)=\int_G d\mu(g)\;\mathrm{exp}\left(\sum_{n=1}^\infty\frac{1}{n}i(p^n,q^n,y^n,z^n)\right),
\end{equation}
\noindent where $d\mu(g)$ is the group measure and the function $i(p,q,y,z)$ denotes the single-particle state index. The latter is determined by representation theory through
\begin{align}
i(p,q&,y,z)=\frac{2pq-p-q}{(1-p)(1-q)}\chi_{adj}(z)\\&+\sum_j\frac{(pq)^{\frac{r_j}{2}}\chi_j(y)\chi_j(z)-(pq)^{\frac{2-r_j}{2}}\overline{\chi}_j(y)\overline{\chi}_j(z)}{(1-p)(1-q)},\nonumber
\end{align}
\noindent where $r_j$ are R-charges, $\chi_{adj}(z)$ is the character of the adjoint representation under which the gauge fields transform, while the second term is a sum over the chiral matter superfields that contains the characters of the corresponding representations of the gauge and flavour groups. In the following, we make use of the fact that SCIs  are identical to particular elliptic hypergeometric integrals.

Define the generalized $A_n$-elliptic hypergeometric integral as
\begin{align}\label{gint}
 &I_{n}^{(m)}(\mathbf{s},\mathbf{t})=\\
&\kappa_{n}\int_{\mathbb{T}^n}
\frac{\prod_{j=1}^{n+1}
\prod_{l=1}^{n+m+2}\Gamma(s_lz_j^{-1},t_lz_j)}{\prod_{1\leq j<k\leq n+1}\Gamma(z_jz_k^{-1},z_j^{-1}z_k)}\; \prod_{k=1}^n
\frac{dz_k}{2\pi i z_k},\nonumber
\end{align}
\noindent with $\prod_{j=1}^{n+1}z_j=1$, $\kappa_n=(p;p)^n(q;q)^n/(n+1)!$, $\mathbf{s}=(s_1,\dots,s_{n+m+2})$, $\mathbf{t}=(t_1,\dots,t_{n+m+2})$, $|s_i|, |t_i|<1$ and the balancing condition $\prod_{i=1}^{n+m+2}s_it_i=(pq)^{m+1}$. The $q$-Pochhammer symbol is defined as $(z;q)_\infty=\prod_{k=0}^\infty(1-zq^k)$, and the elliptic gamma function as

\begin{eqnarray} \label{gamma} &&
\Gamma(z):= \Gamma(z;p,q)
=\prod_{j,k=0}^\infty \frac{1-z^{-1}p^{j+1}q^{k+1}}{1-zp^jq^k},
\\  \nonumber &&
\Gamma(a,b):=\Gamma(a;p,q)\Gamma(b;p,q),
\end{eqnarray}

\noindent for $z\in\mathbb{C}^*$ and $|p|,|q|<1$.  Eq. (\ref{gint}) can be interpreted as the SCI  of an $\mathcal{N}=1$ theory with gauge group $SU(N_c)$ for $N_c=n+1$ and a vector multiplet in its adjoint representation. There is a chiral multiplet in the fundamental and one in the antifundamental of the gauge group, each transforming in the fundamental representation of one of the factors of the flavour group $SU(N_f)\times SU(N_f)$, for $N_f=n+m+2$. Furthermore, there is a global $U(1)_V$ symmetry and the R-symmetry $U(1)_R$. Note that for the sake of brevity, we will not list any R-charges in this paper, as they can be easily recovered from the integral expressions. As shown in \cite{DO}, Seiberg duality is realized by the general
integral identity \cite{rains}
\begin{equation}\label{seiberg}
I_n^{(m)}(\mathbf{s},\mathbf{t})=\prod_{j,k=1}^{n+m+2}\Gamma(t_js_k)I_m^{(n)}(\mathbf{s'},\mathbf{t'})
\end{equation}
\noindent with the arguments $\mathbf{s'}=(S^{\frac{1}{m+1}}/s_1,\dots,S^{\frac{1}{m+1}}/s_{n+m+2})$ and $\mathbf{t'}=(T^{\frac{1}{m+1}}/t_1,\dots,T^{\frac{1}{m+1}}/t_{n+m+2})$, where $S=\prod_{j=1}^{n+m+2}s_j$, $T=\prod_{j=1}^{n+m+2}t_j$, $ST=(pq)^{m+1}$ and $|t_k|,|s_k|,|S^{\frac{1}{m+1}}/s_k|,|T^{\frac{1}{m+1}}/t_k|<1$. The operation $n\leftrightarrow m$ gives the correct dual symmetry groups since $N_f=n+m+2\rightarrow N_f$ and $N_c=n+1\rightarrow m+1=N_f-N_c$.

For $m=0$, Eq. (\ref{seiberg}) reduces to the exact evaluation formula \cite{spi,spi03}
\begin{equation}\label{aint}
I_n^{(0)}(\mathbf{s},\mathbf{t})=\prod_{k=1}^{n+2}\Gamma\Big(\frac{S}{s_k},\frac{T}{t_k}\Big)
\prod_{k,l=1}^{n+2}\Gamma(s_kt_l).
\end{equation}

\noindent This is an example of $s$-confinement \cite{CSS}: the infrared is described only by gauge-invariant operators, and the origin of the classical moduli space remains a vacuum even after quantizing the theory (chiral symmetry is intact). Furthermore, a confining superpotential is generated dynamically.

We define \cite{spi2004,SW} as a Bailey pair with respect to the parameter $t$ a pair of functions $\alpha(z,t)$ and $\beta(w,t)$ satisfying the relation $\beta(w,t)=M(t)_{wz}\alpha(z,t)$, where $M(t)_{wz}$ is an elliptic hypergeometric integral operator. The (integral) Bailey lemma states that given such a pair of functions, one automatically obtains another Bailey pair with respect to a new parameter $st$, i.e. $\beta'(w,st)=M(st)_{wz}\alpha'(z,st)$. This pair is related to the original one by the transformations $\alpha'(w,st)=D(s,t^{-\frac{n-1}{2}}u)_w\alpha(w,t)$ and
$\beta'(w,st)=D(t^{-1},s^{\frac{n-1}{2}}u)_wM(s)_{wz}D(ts,u)_z\beta(z,t)$, where $D(t,u)_z$ is a function with the property $D(t,u)_z D(t^{-1},u)_z=1$ and $u$ is a new arbitrary parameter. From these expressions, it is easy to derive the star-triangle relation
\begin{align}\label{startriangle}
&M(s)_{wz}D(st,u)_zM(t)_{zx}=\\
&D(t,s^{\frac{n-1}{2}}u)_wM(st)_{wx}D(s,t^{-\frac{n-1}{2}}u)_x.\nonumber
\end{align}
\noindent Repeated application of the Bailey lemma leads to infinite recursion relations referred to as Bailey chains. The $A_n$ version of the Bailey lemma is obtained by identifying Eq. (\ref{startriangle}) with Eq. (\ref{aint}), which leads to
\begin{align}\label{mop}
&M(t)_{wz}f(z):=\\&\kappa_n \int_{\mathbb{T}^n}
\frac{\prod_{j,k=1}^{n+1}\Gamma(tw_jz_k^{-1})f(z)}
{\Gamma(t^{n+1})\prod_{1\leq j<k\leq n+1}\Gamma(z_jz^{-1}_k,
z_j^{-1}z_k)}\prod_{k=1}^{n}\frac{dz_k}{2\pi\textup{i}z_k},\nonumber
\end{align}
\noindent and
\begin{equation}
D(t,u)_z:=\prod_{j=1}^{n+1}\Gamma(\sqrt{pq}t^{-\frac{n+1}{2}}\frac{u}{z_j},
\sqrt{pq}t^{-\frac{n+1}{2}}\frac{z_j}{u}).
\end{equation}
\noindent The operator $M(t)_{wz}$ was first defined for $n=1$ in \cite{spi2004} and for arbitrary $n$ in \cite{SW}. For certain constraints on $t$ and $w_j$ it satisfies the Fourier type inversion
relation, $M(t)_{wz}^{-1}=M(t^{-1})_{wz}$.

This identification of operators leads to many interesting nontrivial relations, e.g. to
the recursion formula \cite{spi}
\begin{equation}\label{equiver}
I_{n}^{(m+1)}(\mathbf{s},\mathbf{t})=\mathbf{Q}_n^mI_n^{(m)}(\tilde{\mathbf{s}},\mathbf{t}),
\end{equation}
\noindent where $\mathbf{Q}_n^m$ is the integral operator
\begin{align}
&\mathbf{Q}_n^mf(w):=\zeta(v)\times \\
&\int_{\mathbb{T}^n}
\frac{\prod_{j=1}^{n+1}
\Gamma(\frac{t_{n+m+3}w_j}{v^n})\prod_{l=1}^{n+2} \Gamma(\frac{s_l}{vw_j})}{\prod_{1\leq j<k\leq n+1}\Gamma(w_jw_k^{-1},w_j^{-1}w_k)}f(w)\prod_{k=1}^n \frac{dw_k}{2\pi\textup{i}w_k},\nonumber\end{align}
\noindent with $\tilde{\mathbf{s}}=(vw_1,\ldots,vw_{n+1},s_{n+3},\ldots,s_{n+m+3})$ and
\begin{equation}\label{zeta}
\zeta(v)=\frac{\kappa_{n}}{\Gamma(v^{n+1})}\prod_{l=1}^{n+2}\frac{\Gamma(t_{n+m+3}s_l)}{\Gamma(v^{-n-1}t_{n+m+3}s_l)}.
\end{equation}
\noindent The parameter $v$ is related to $t_{n+m+3}$ by $v^{n+1}=t_{n+m+3}(pq)^{-1}\prod_{i=1}^{n+2}s_i$.

Eq. (\ref{equiver}) can be understood as an algorithm for constructing the SCI  of a linear $\mathcal{N}=1$ quiver gauge theory. To see this, consider Eq. (\ref{gint}) with $m=0$, which is the SCI  of an $\mathcal{N}=1$ theory with gauge group $SU(n+1)$, as can be read off from the denominator of the integrand and the fact that there is just an integral over one set of variables $z_j$ satisfying $\prod_{j=1}^{n+1}z_j=1$. Applying the operator $\mathbf{Q}_n^0$ to this expression adds another $SU(n+1)$ gauge group to the theory. There is now a chiral multiplet transforming in the fundamental representation of the new, and in the antifundamental of the original gauge group, as expected from a quiver. This procedure can be iterated indefinitely, yielding a linear quiver of arbitrary length. In addition to the fields mentioned above, we also get an additional chiral multiplet transforming in the fundamental representation of the new gauge factor. It is important to note that while the flavour symmetry on the left hand side is $SU(N_f+1)\times SU(N_f+1)$, the full flavour symmetries of the quiver on the right hand side of Eq. (\ref{equiver}) are given by $SU(N_f-N_c)\times SU(N_c+1)\times U(1)$ and $SU(N_f)\times U(1)$, subgroups of the flavour symmetry group of SQCD on the left hand side. As can be read off from its definition, fields charged under the $SU(N_c+1)$ factor are part of the $\mathbf{Q}_n^m$ operator, while $SU(N_f-N_c)$ and $SU(N_f)$ arise directly from $I_n^{(m)}$. The flavour symmetry of the latter is broken by the replacement of the parameters $\mathbf{s}$ by $\tilde{\mathbf{s}}$. This mismatch in symmetries points towards symmetry enhancement in the IR. Furthermore, the duality expressed by Eq. (\ref{equiver}) is a realization of $s$-confinement: as one can see from counting the number of flavours attached to each node, one of the nodes $s$-confines. We will elaborate on this in \cite{BS}, where we will also study the field content in more detail. 

A surprising observation is that no matter how long the quiver one has generated with the help of Eq. (\ref{equiver}) is, it can be rewritten in terms of a single integral through Eq. (\ref{gint}). Given that all of the integrals generated by the Bailey lemma can be interpreted as SCIs, this leads us to the conjecture that the electric part of SQCD, with its SCI  given by Eq. (\ref{gint}), has a large number of dual linear quivers, related to the original theory by $s$-confinement and symmetry enhancement. To see how many, simply count the number of possible starting points of the iteration, the result is $m$, which can be rewritten as $N_f-N_c-1$. Applying Seiberg duality to the resulting quivers adds even more dual theories. One possible equation arising from this would be

\begin{align}\label{chain}
I_n^{(m)}(\mathbf{s},\mathbf{t})&=  \mathbf{Q}_n^{m-1} \cdots  \mathbf{Q}_n^{i} I_n^{(i)}(\tilde{\mathbf{s}},\mathbf{t})\\
&=\mathbf{Q}_n^{m-1} \cdots  \mathbf{Q}_n^{i} c_i^n(\tilde{\mathbf{s}},\mathbf{t})    \mathbf{Q}_i^{n-1} \cdots   \mathbf{Q}_i^{j} I_i^{(j)}(\tilde{\mathbf{s}}',\mathbf{t}')\nonumber,
\end{align}

\noindent where $i=0,\ldots, m-1$ and $j=0,\ldots,n-1$ denote starting points
of the iteration and enumerate possible quivers. Seiberg duality is realized through  $I_n^{(i)}(\tilde{\mathbf{s}},\mathbf{t})=c_i^n(\tilde{\mathbf{s}},\mathbf{t}) I_i^{(n)}(\tilde{\mathbf{s}}',\mathbf{t}')$, where the coefficient $c_i^n(\tilde{\mathbf{s}},\mathbf{t})$ corresponds to that in Eq. (\ref{seiberg}). Evidently, more than one coefficient of this type can show up in expressions like those of Eq. (\ref{chain}). In principle, each application of Eq. (\ref{seiberg}) and Eq. (\ref{equiver}) adds an additional tilde or prime to the parameters, but we try to keep the notation simple by not writing them explicitly.

As an example, consider Eq. (\ref{gint}) with $n=m=2$, which corresponds to $N_c=3$ and $N_f=6$. One can either start with $m=0$, and iterate Eq. (\ref{equiver}) twice, or start with $m=1$ and apply it once, to end up at $I_2^{(2)}$. The result is that we have two different indices of quiver gauge theories that are equivalent to the electric indices of SQCD for our choice of colours and flavours. This is shown in Fig. \ref{network}, the $Q$-operation relates the theory on top with a single gauge group to two quivers. We can now apply Seiberg duality of Eq. (\ref{seiberg}) to the original integral in both expressions, this is denoted as $S$ in Fig. \ref{network}. One of the resulting theories can now again be rewritten through Eq. (\ref{equiver}) in an analogous manner. The whole logic can be applied once again, until no more possibilities arise. The complete web of dualities that arises from this is shown in Fig. \ref{network}. We get a total of 
ten distinct theories whose SCIs  match, indicating a duality
relation. The same procedure can be applied to the magnetic theory as well, leading to, in this example, another set of ten dual theories. For a general number of colours and flavours, the result will not be symmetric. Another noteworthy aspect of this new duality web is the fact that the SCI  of SQCD with arbitrary flavours can be generated from the $s$-confining theory with $m=0$ ($N_c=n+1$ and $N_f=N_c+1$) by quivering it, i.e. by repeated application of Eq. (\ref{equiver}).

The quivers generated by Eq. (\ref{equiver}) are free of gauge anomalies. Consider a node corresponding to a vector multiplet. Oriented edges connecting it to adjacent nodes correspond to bifundamental fields transforming both under the gauge symmetry and another gauge or flavour symmetry. The original gauge symmetry is anomaly free if the weighted sum over the ranks of the adjacent symmetry groups ($1$ for the fundamental, $-1$ for the antifundamental representation) vanishes. This is the case for the quiver on the right hand side of Eq.(\ref{equiver}) and subsequently for all quivers in the duality web. 

We have also checked the matching of global anomalies by computing the triangle diagrams of the global symmetries of the quiver, including the $U(1)_R$ symmetry. All anomaly coefficients, which will be presented explicitly in \cite{BS}, match with those of the corresponding subgroups of the index on the left hand side of Eq. (\ref{equiver}). In \cite{SV2} it was shown that $SL(3;\mathbb{Z})$ transformation properties of the elliptic 
hypergeometric integrals describe 't Hooft anomaly
matching conditions. In \cite{Razamat}, modular transformations where studied in the Schur limit of the index, where the $SL(3;\mathbb{Z})$ group 
reduces to $SL(2;\mathbb{Z})$. In the present context
anomaly matching means that in relation (11) the sum 
of Bernoulli polynomials $B_{33}(u;\mathbb{\omega})$ 
(with appropriate arguments) associated with a transformation of the integral
operator $Q_n^m$ (there is also an addition coming from
 the modular transformation of the multiplier $\kappa_n$) and of that for the integral (SCI) 
$I_n^{(m)}$ is equal to the corresponding
polynomial for the integral (SCI) $I_n^{(m+1)}$.
This picture should agree also with the computation
of partition functions for our quiver theories along
the lines of \cite{ACM,ACDKLM,BBK}, which are proportional 
to SCIs up to an exponential of the Casimir energies.

At this point, we would like to make a few comments on the physical interpretation of the Bailey lemma, especially the integral operator $M(t)_{wz}$ of Eq. (\ref{mop}). In \cite{DS}, it was considered for $n=1$, where its connection to the Sklyanin algebra, a particular realization of a quantum algebra related to the Yang-Baxter equation was discussed. An interesting aspect of this work is the emergence of a recurrence relation for the intertwining operator $M(t)_{wz}$ that involves a specific finite difference operator composed of Jacobi theta functions. Similar structures arise in the study of generalized $\mathcal{N}=2$ quiver gauge theories of class $\mathcal{S}$ \cite{Gaiotto}. It was shown \cite{GRR} that the SCI  associated with an IR theory defined on a Riemann surface, notably in the presence of surface defects, is determined by the pole structure of the index of the corresponding UV theory. To be more precise, one has to calculate the residues corresponding to poles of elliptic gamma functions
appearing in the integrand of the UV index. This is done with the help of a difference operator similar to the one derived in  \cite{DS} from residue calculus.
Note that $M(t)$ is converted into a finite-difference operator
corresponding to defect insertions by restricting to $t^{n+1}=q^{-r}p^{-s}$
with $r,s\in \mathbb{Z}_{\geq 0}$. For generic $t$, the integral operator $M(t)$ is a very
general object describing an insertion of a whole nontrivial interacting
field theory with gauge and matter fields.

It would be interesting to study the new duality web from the point of view of a six-dimensional construction and see whether the intertwining operator and the related elements of the Sklyanin algebra can be connected to $\mathcal{N}=1$ analogues of known $\mathcal{N}=2$ surface defects. It is also important to clarify the relation to the $\mathcal{N}=1$ linear quivers of \cite{BB} and to the  constructions of \cite{Yagi} and \cite{HM}. In \cite{Yagi}, the author considered brane box models giving rise to $\mathcal{N}=1$ quiver gauge theories and related the SCI  to the correlation function of line operators in a two-dimensional topological QFT on a torus, which in turn can be related to two-dimensional lattice models as it was observed first in \cite{conm}.

Finally, let us mention that the described duality web is not the only interesting
structure that arises from the $A_n$-Bailey lemma. We have limited ourselves to
a particular class of linear quivers in this article, whereas relation (\ref{equiver}) shows that the full web of dualities contains an even larger set of theories. To see this, note that the ``quivering'' operator $\mathbf{Q}_n^m$
only acts on a subset of parameters, i.e. $s_i$ for $i= 1,\ldots, n+1$.
However, action on the remaining flavour parameters is legitimate,
and combining operators that act on different sets of parameters
leads to a substantially more complicated duality web. In terms of Fig. \ref{network}, this means that to each bubble corresponding to a linear quiver, we actually have
to attach a larger number of dual theories. Furthermore, the set of dualities described in Sect. 11.2 of the first
paper in \cite{SV1} is a part of this web. We will present
a detailed exposition of the $A_n$-Bailey lemma consequences for constructing
dual field theories in an upcoming paper \cite{BS}.

After completion of this paper, we were informed about the work of \cite{MY}, where the relation
of the Sklyanin algebra to the insertion of surface defects in the context of
six-dimensional theories is discussed.

\bigskip
F.B. would like to thank A.I. Soloviev and C. Ecker, I. Garc\'{i}a-Etxebarria and Timm Wrase for helpful discussions. F.B. was supported by the Austrian Science Fund FWF, project no.  P26366, and the FWF doctoral program Particles \& Interactions, project no. W1252. V.S. was supported by the Russian Foundation for Basic Research grant no. 16-01-00562.

\end{document}